\documentclass[a4paper,11pt]{article}
\usepackage{jinstpub}
\usepackage{hyperref}
\usepackage{orcidlink}
\usepackage{comment}
\usepackage{caption}
\usepackage[toc,page]{appendix}
\usepackage{csquotes}
\usepackage{subcaption}
\usepackage{amsmath}
\usepackage{graphicx}
\usepackage{makecell}
\usepackage{enumitem}
\usepackage{float}

\title{Characterization and Monte Carlo Simulation of Wavelength-Shifting Absorption Length in Plastic Optical Fibers}

\author[a,{*\dag}]{W.~Bae\orcidlink{0000-0002-7646-7577}\note[{*}]{Corresponding author.}\note[{\textdagger}]{Present address: Department of Physics and Astronomy, Northwestern University, Evanston, IL 60208, USA.},}
\author[a]{L.~Baker\orcidlink{0009-0006-5512-642X},}
\author[a]{K.~Chen\orcidlink{0009-0008-5642-7624},}
\author[a]{J.~Cho\orcidlink{0009-0004-7530-7231},}
\author[a]{K.~Lang\orcidlink{0000-0003-1269-7223},}
\author[a,*]{C.~Lee\orcidlink{0009-0003-5915-3642},}
\author[a]{E.~Liang\orcidlink{0009-0008-6995-3842},}
\author[a]{C.~Mathurin\orcidlink{0009-0004-0438-3640},}
\author[a]{C.~Murthy\orcidlink{0000-0001-9044-7946},}
\author[a]{D.~Myers\orcidlink{0000-0001-8402-7240},}
\author[a]{S.~Nguyen\orcidlink{0009-0009-9622-6429},}
\author[a]{M.~Proga\orcidlink{0000-0002-0303-5159},}
\author[a]{M.~Zalikha\orcidlink{0009-0002-7045-6022},}
\author[a]{J.~Zey\orcidlink{0009-0008-3005-6642}}
\affiliation[a]{Department of Physics, University of Texas at Austin,\\ 
1 University Station, Austin, TX 78712-0264, USA}
\emailAdd{wonseokb@utexas.edu}

\abstract{
We report an experimental characterization of the wavelength-shifting (WLS) absorption length of the EJ-160I fiber core material from Eljen Technology.
Fiber-core samples of different thicknesses were employed, and the transmitted intensity of incident LED light was measured to extract the WLS absorption length at 405.7\,nm.
The resulting value, \((0.823 \pm 0.038)\,\mathrm{mm}\), was used to normalize the WLS absorption length spectrum derived from the absorption spectrum of the core material.
An independent measurement at 392.5\,nm yielded a WLS absorption length of \((1.404 \pm 0.074)\,\mathrm{mm}\), differing from the value predicted by the reconstructed WLS absorption length spectrum by 3.16\%, corresponding to \(0.43\sigma\).
The resulting spectrum was implemented as an optical input in a Geant4-based optical Monte Carlo simulation of the EJ-160I fiber to examine the role of the WLS absorption length in light output, light attenuation, and the emission spectrum.
}


\begin{document}
\maketitle
\flushbottom


\section{Introduction}
\label{sec:Introduction}

\begin{figure}[h!]
    \centering
    \includegraphics[width=0.9\linewidth]{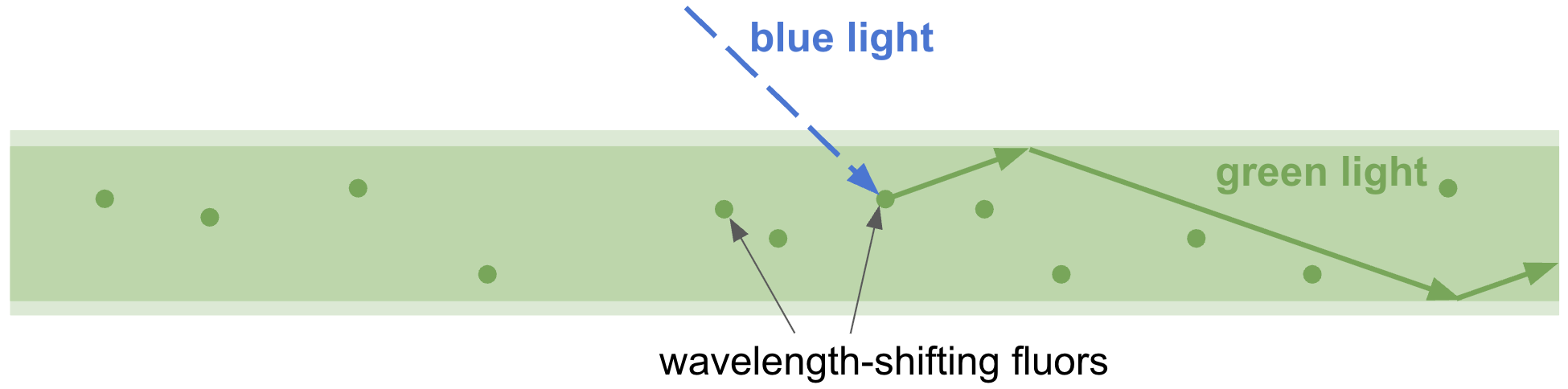}
    \caption{Conceptual schematic illustration of light collection in a wavelength-shifting (WLS) fiber. WLS fluors in the fiber core absorb incident photons and re-emit light at longer wavelengths, and the shifted light is guided along the fiber by total internal reflection.
    This figure is adapted from~\cite{Bae_PhDthesis}.
    }
    \label{fig:WLS_fiber_mechanism}
\end{figure}

Plastic wavelength-shifting (WLS) fibers consist of a fluorescently doped polymer core surrounded by a lower-index cladding.
As illustrated in figure~\ref{fig:WLS_fiber_mechanism}, WLS fluors within the fiber core absorb incident photons and re-emit light at longer wavelengths. The wavelength-shifted photons are guided along the fiber by total internal reflection~\cite{Ruchti:1996,Pla-Dalmau:2001,Achenbach:fiber-cladding_light}.
Owing to this combination of wavelength conversion and light guiding, WLS fibers have found broad use in particle- and nuclear-physics detectors, including MINOS~\cite{MINOS-Michael:2008bc,Avvakumov:2005ww}, NOvA~\cite{Ayres:2004js}, GERDA~\cite{Ackermann:2012xja}, and LEGEND-200~\cite{LEGEND:2025insdet}.

There are two characteristic length scales associated with absorption in a WLS fiber: the attenuation length and the wavelength-shifting absorption length.
The attenuation length, on the meter scale, describes the decrease in guided light intensity as it propagates along the fiber~\cite{Amos1990:fiber-cladding_light,Drexlin:1995,Papandreou:2009}.
The WLS absorption length \(L_{\mathrm{WLS}}\), generally on the millimeter scale, is the characteristic distance over which incident photons are absorbed by wavelength-shifting fluors prior to re-emission.
Each characteristic length is the distance at which the relevant light intensity falls to \(1/e\) of its initial intensity.
The two optical properties are required as independent input parameters, \texttt{ABSLENGTH} and \texttt{WLSABSLENGTH}, in Geant4 optical Monte Carlo simulations of WLS fibers~\cite{geant4,Geant4:2006,Geant4:2016}.

Meter-scale attenuation lengths are commonly available from manufacturer specifications or standard fiber-characterization measurements~\cite{bae2025optical, bae2025response, bae2026bcf20xl}, whereas the corresponding WLS absorption length is less readily available as a separate optical input.
This work presents a transmission-based method for measuring the WLS absorption length at discrete wavelengths and using these measurements to reconstruct the corresponding WLS absorption length spectrum for implementation in optical simulations.

\section{Measurement Setup}

\begin{figure}[h!]
  \centering
    \includegraphics[width=0.9\textwidth]{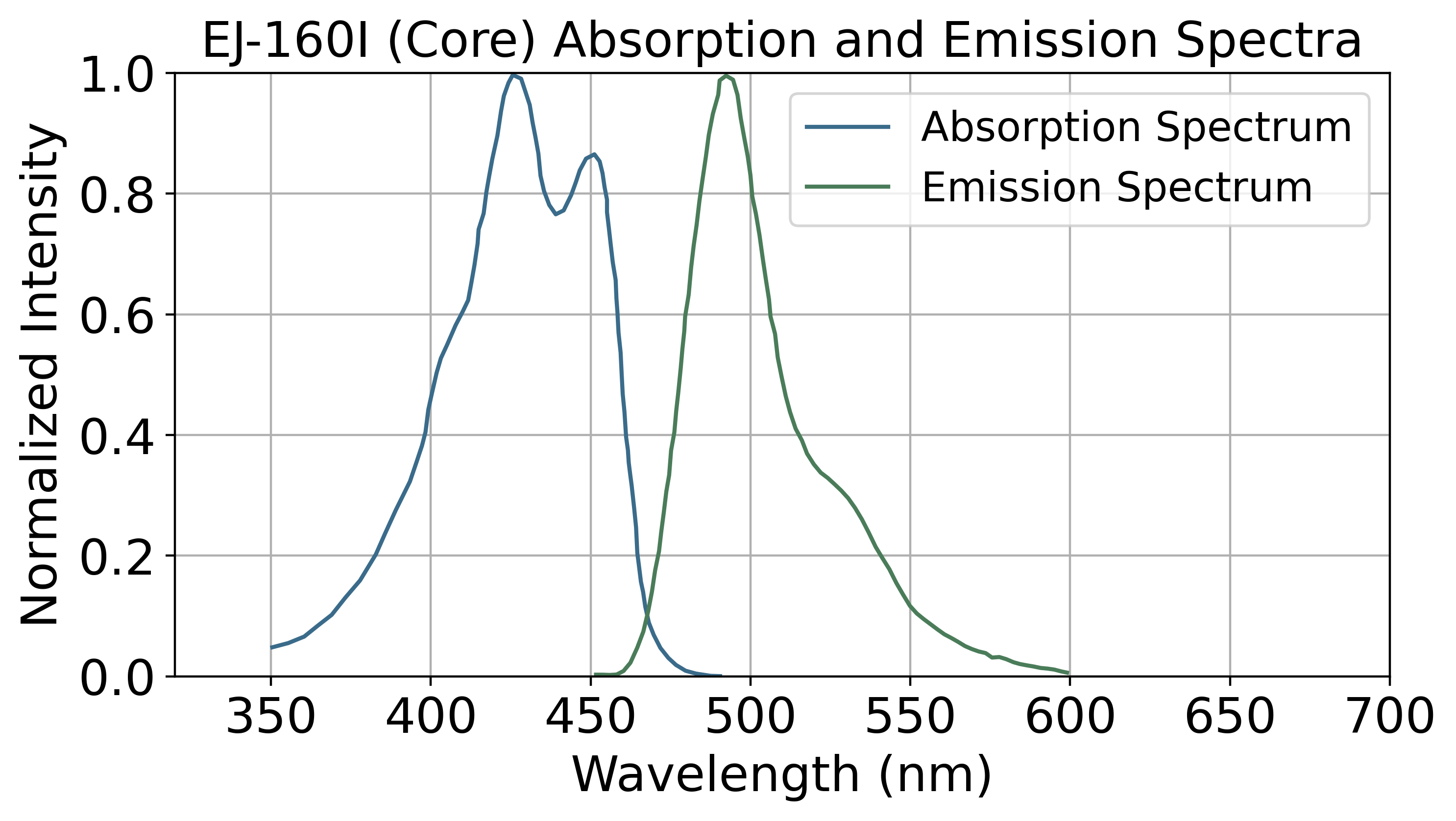}

  \caption{Normalized absorption spectrum (blue) and emission spectrum (green) of the EJ-160I fiber core material, based on manufacturer data~\cite{eljen}.}
  \label{fig:spectra}
\end{figure}

In this work, the WLS absorption length is characterized from the transmitted intensity of LED light through EJ-160I fiber-core samples of different thicknesses.
The optical absorption and emission spectra of the EJ-160I fiber core material are shown in figure~\ref{fig:spectra}.
The EJ-160I fiber consists of a polystyrene core doped with WLS fluors and surrounded by a thin polymethyl methacrylate (PMMA) cladding; the fiber-core samples were used to characterize the optical absorption properties of the core material.
Four samples of different thicknesses were prepared. The two thinnest samples, 0.787\,mm and 0.864\,mm, were single fiber cores, while the remaining two samples, 1.630\,mm and 2.464\,mm, were obtained by stacking two and three fiber cores side by side, respectively.

\noindent
\begin{figure}[h!]
    \centering
    \includegraphics[width=0.9\linewidth]{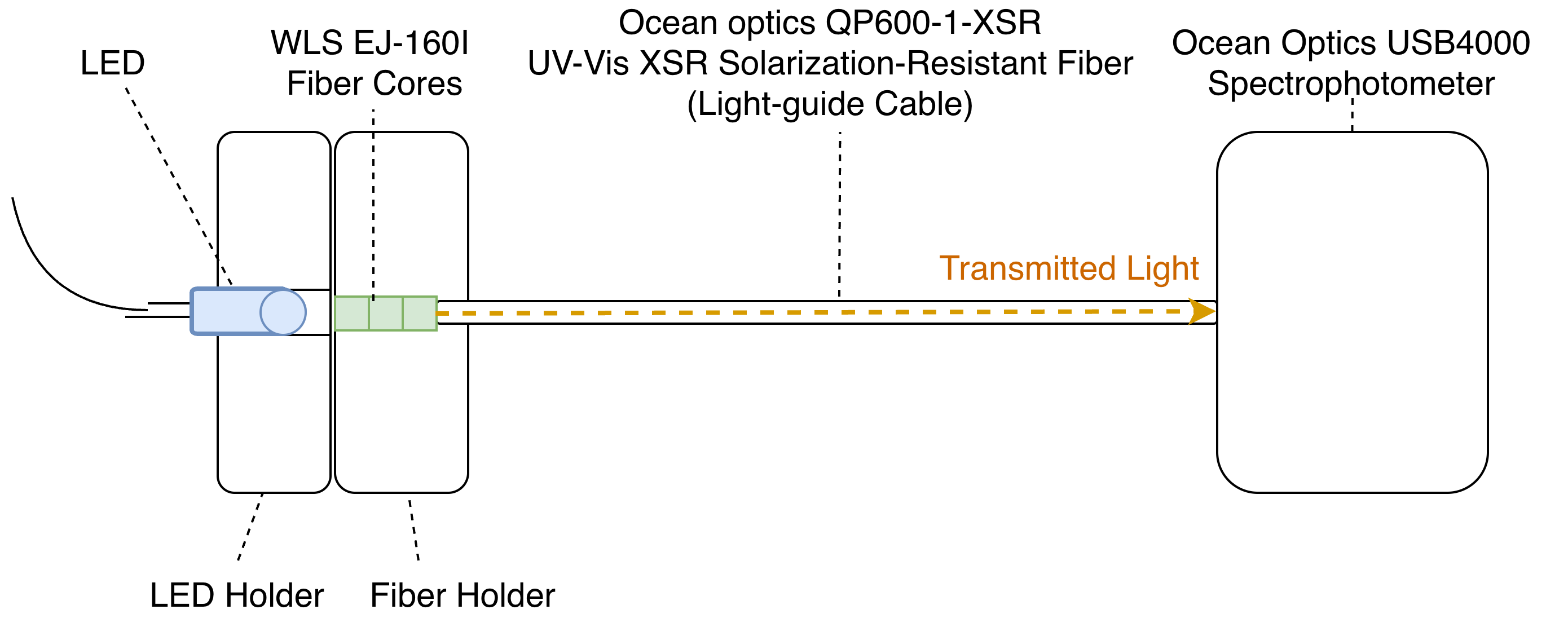}
    \caption{Schematic illustration of the experimental setup for wavelength-shifting absorption length measurement.}
    \label{fig:measurementsetup}
\end{figure}

\begin{figure}[h!]
    \centering
    \includegraphics[width=0.8\linewidth]{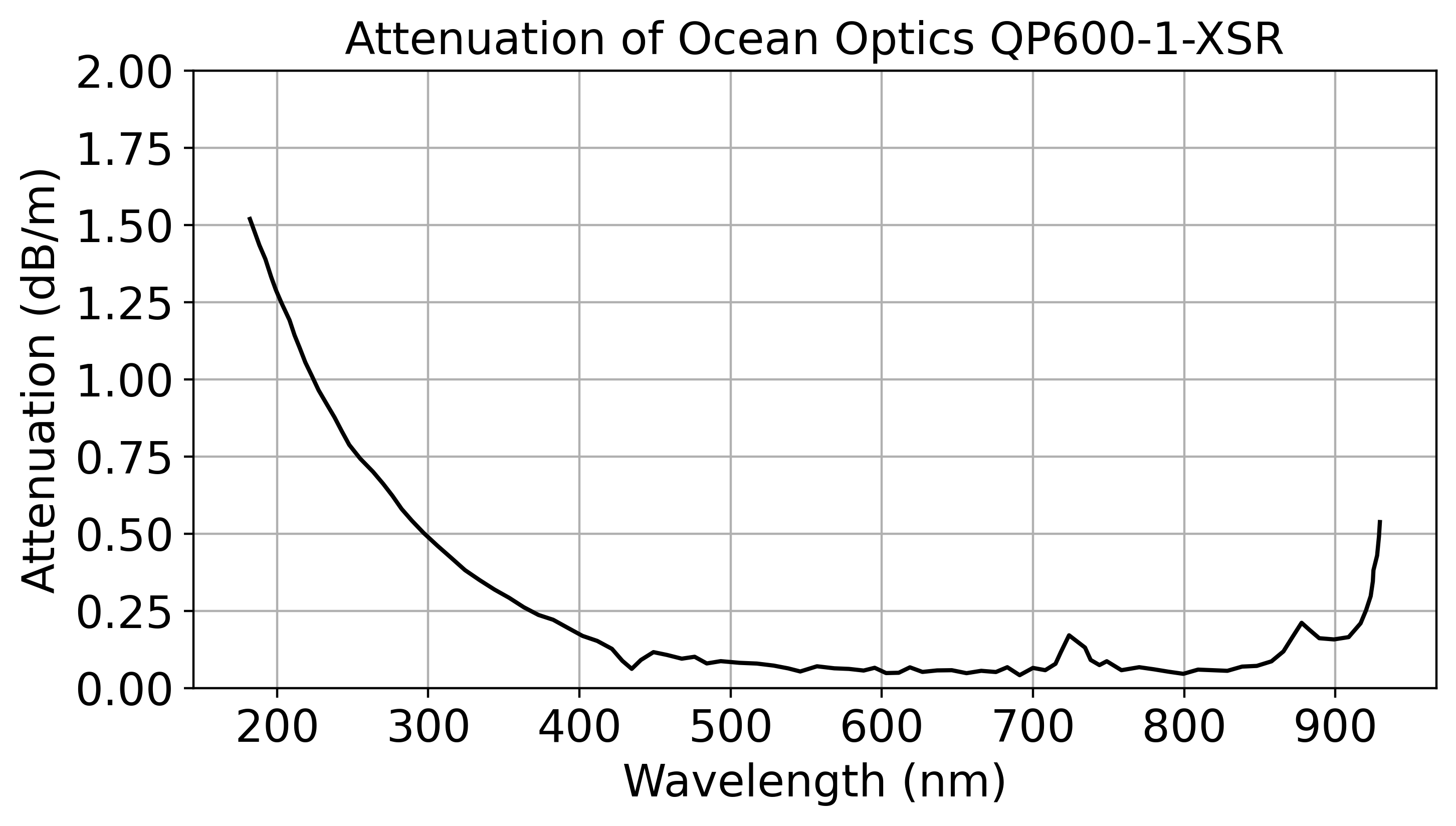}
    \caption{Spectral attenuation of the Ocean Optics QP600-1-XSR light-guide cable used to couple the fiber-core samples to the spectrophotometer, based on manufacturer data~\cite{Ocean-Optics}. Over the 1\,m cable length, the corresponding transmission is 96.3\% at 405.7\,nm and 95.6\% at 392.5\,nm.}
    \label{fig:xsr}
\end{figure}

The experimental setup is illustrated in
figure~\ref{fig:measurementsetup}.
The LED was mounted in an LED holder, while the fiber-core sample was secured in a grooved fiber holder and placed in direct contact with the end face of a light-guide cable~\cite{Ocean-Optics}.
The opposite end of the light-guide cable was coupled to the spectrophotometer~\cite{Ocean-Optics}, which covers the near-ultraviolet and visible spectral regions (approximately 350--800\,nm) with a wavelength precision of approximately 0.21\,nm.
The light-guide cable was an Ocean Optics QP600-1-XSR~\cite{Ocean-Optics}, a solarization-resistant UV-Vis fiber with a 600\,$\mu$m core and a length of 1\,m.
Since all recorded spectra correspond to light transmitted through this cable, they were corrected for its spectral attenuation, shown in figure~\ref{fig:xsr}.

Two LEDs were used, with peak emission near 405.7\,nm and 392.5\,nm. Both were driven at 120\,MHz, with amplitudes set to 3.8\,V for the 405.7\,nm LED and 2.9\,V for the 392.5\,nm LED.
Each spectrum was generated by averaging 10 consecutive acquisitions, with an integration time of 60\,$\mu$s for the 405.7\,nm LED and 550\,$\mu$s for the 392.5\,nm LED, yielding stable spectral profiles of transmitted light.
All measurements were conducted in a dark environment, and each recorded spectrum was corrected by subtracting the residual background spectrum.

Five measurement configurations were used: an LED-only reference and four transmitted-light configurations with fiber-core thicknesses of 0.787\,mm, 0.864\,mm, 1.630\,mm, and 2.464\,mm.
For each configuration, three spectra were acquired, with the fiber-core samples re-mounted between repeated acquisitions to account for measurement-to-measurement variations associated with alignment.

\section{Wavelength-Shifting Absorption Length}
\label{sec:wls_absorption_length}

\subsection{WLS absorption length at 405.7 nm}
\label{sec:wls_absorption_405}

\begin{figure}[h!]
    \centering
    \includegraphics[width=0.8\linewidth]{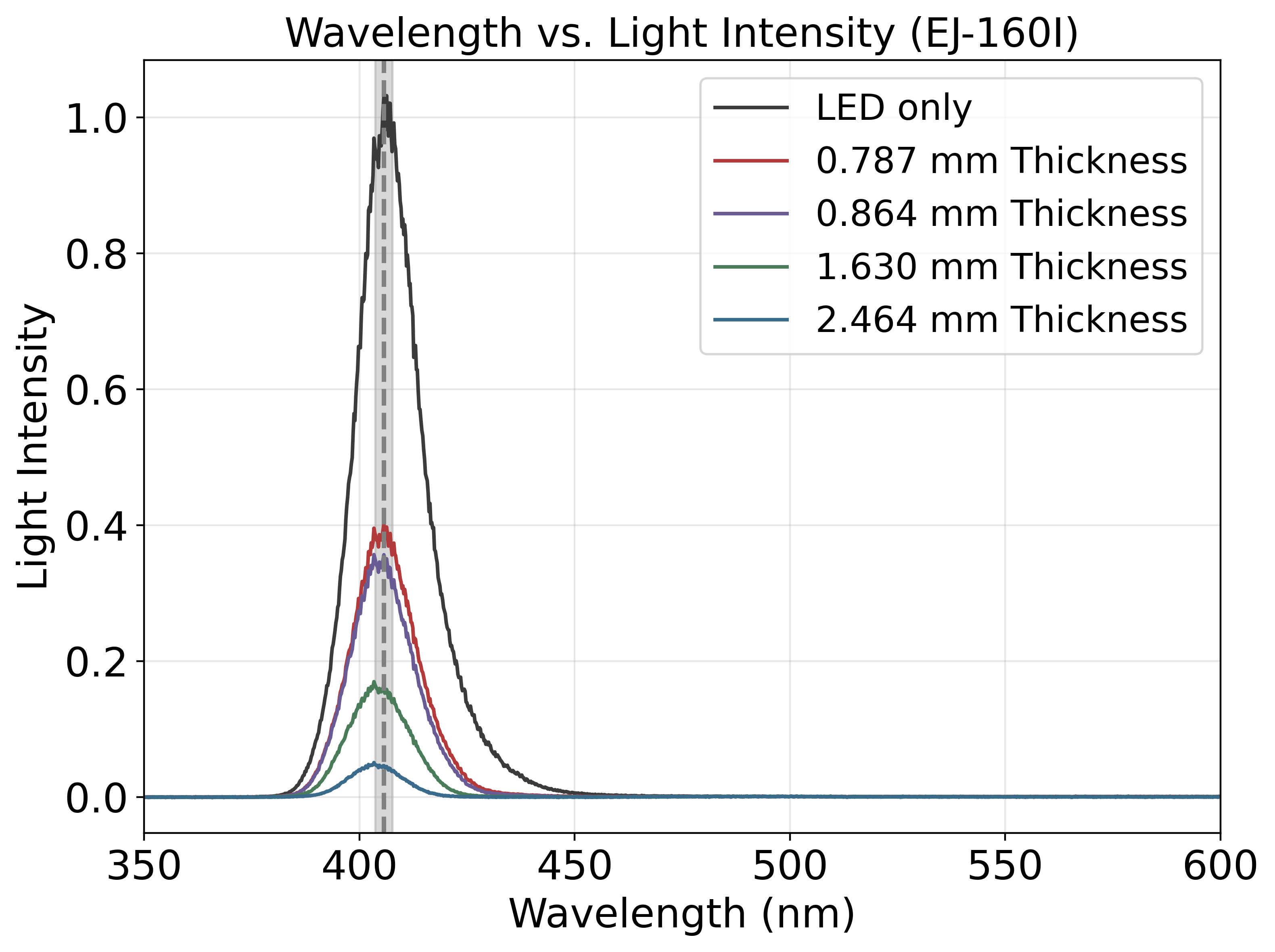}
    \caption{Normalized incident LED intensity and transmitted-light spectra through fiber-core samples with four different thicknesses, corrected for the transmission of the light-guide cable (figure~\ref{fig:xsr}). The shaded gray band indicates the 403.7--407.7~nm window centered on the LED emission peak near 405.7~nm.}
    
    \label{fig:405_Spectrum}
\end{figure}

The transmitted-light spectra obtained using the LED with an emission peak near 405.7\,nm are shown in figure~\ref{fig:405_Spectrum}.
For each configuration, the transmitted-light intensity was extracted at 405.7\,nm and used to characterize the WLS absorption length at $\lambda = 405.7\,\mathrm{nm}$.








Because the sample thicknesses are on the millimeter scale, the contribution from meter-scale propagation losses is expected to be minimal relative to WLS absorption.
At a fixed wavelength, the measured transmission ratio was therefore modeled using the Beer--Lambert relation with an effective optical-coupling factor~\cite{Swinehart:1962,Li:2014RTV}:
\begin{equation}
    \frac{I}{I_0}
    =
    A\exp\!\left(-\frac{x}{L_{\mathrm{WLS}}}\right),
    \qquad x>0,
    \label{eq:1}
\end{equation}
where \(I_0\) is the intensity measured in the LED-only reference configuration, \(I\) is the transmitted intensity measured with a fiber-core sample, and \(A\) ($0 < A \le 1$) is a factor accounting for the optical coupling introduced at the LED--sample and sample--light-guide-cable interfaces, with values closer to 1 representing more efficient light transmission with minimal coupling loss.
For the stacked configurations, adjacent fiber-core samples were pressed into direct contact to minimize interfacial air gaps.
Each stack was therefore treated as a single effective sample with a thickness equal to the sum of the individual core thicknesses.

Taking the logarithm of eq.~\eqref{eq:1} gives
\begin{equation}
    \ln\!\left(\frac{I_0}{I}\right)
    =
    \frac{x}{L_{\mathrm{WLS}}}-\ln A,
    \qquad x>0.
    \label{eq:2}
\end{equation}
Accordingly, \(\ln(I_0/I)\) is expected to scale linearly with the total fiber-core thickness \(x\), and \(L_{\mathrm{WLS}}\) is obtained as the inverse of the fitted slope.

\begin{figure}[h!]
    \centering
    \includegraphics[width=0.85\linewidth]{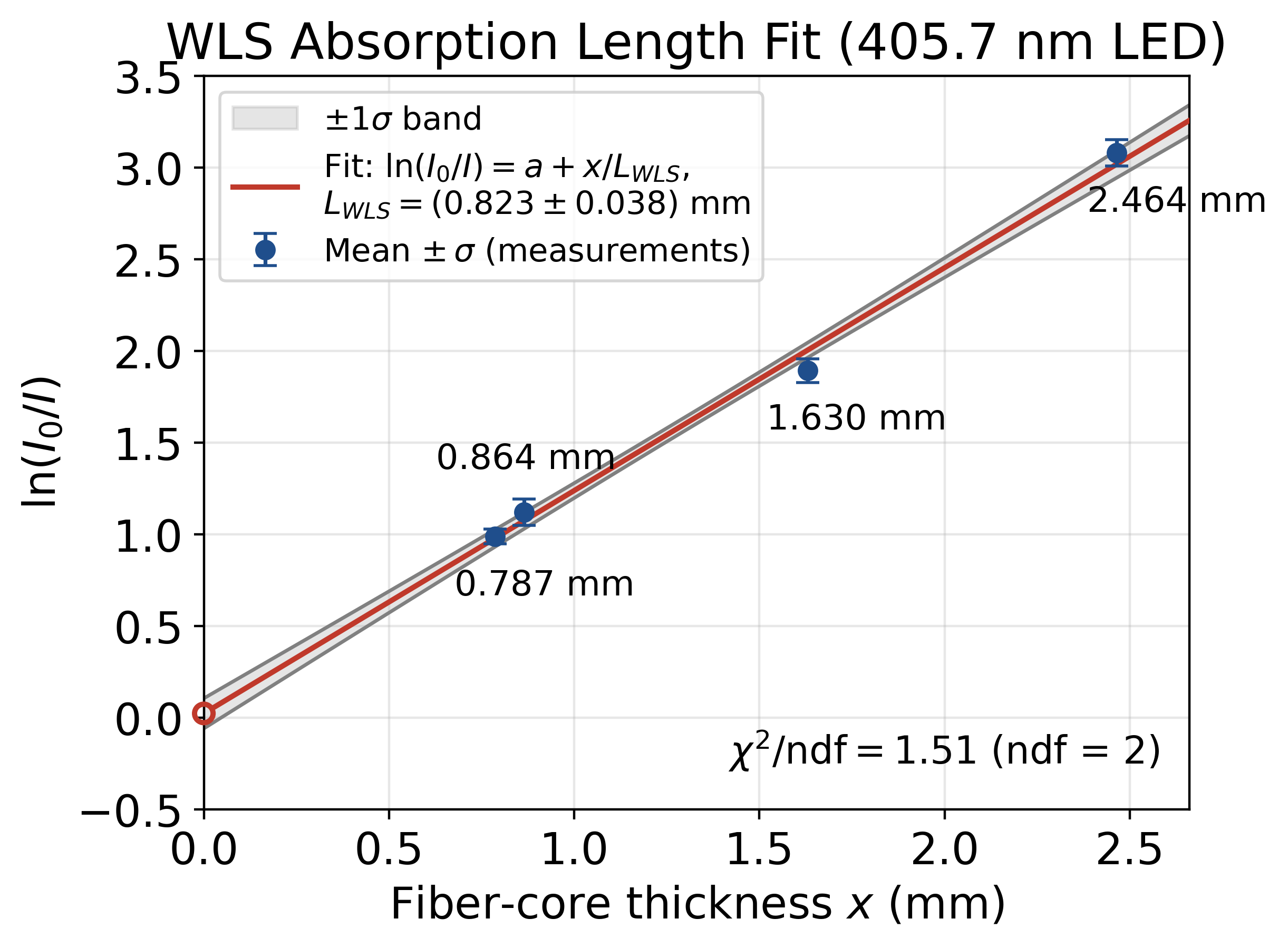}
    \caption{Extraction of the WLS absorption length at 405.7~nm from the thickness dependence of light attenuation. The intensity ratio $\ln(I_0/I)$ is plotted against fiber-core thickness $x$ for the 405.7~nm LED. The solid red line shows the linear fit $\ln(I_0/I) = a + x/L_{WLS}$ for $x > 0$, yielding $L_{WLS} = (0.823 \pm 0.038)$\,mm. The shaded gray band denotes the corresponding $\pm 1\sigma$ uncertainty band of the fit.}
    \label{fig:405_Linear}
\end{figure}

The linear relation between $\ln(I_0/I)$ and the fiber-core thickness is shown in figure~\ref{fig:405_Linear}, where the error bars represent the standard deviation of the extracted $\ln(I_0/I)$ values from the three repeated measurements for each thickness configuration.
From eq.~\eqref{eq:2}, the inverse of the fitted slope gives the WLS absorption length, yielding $L_{\text{WLS}}(\lambda = 405.7\,\mathrm{nm}) = (0.823 \pm 0.038)\,\mathrm{mm}$ for the EJ-160I fiber. The corresponding fitted optical-coupling factor is $A = 0.973 \pm 0.061$.

%
%

\subsection{WLS absorption length spectrum}
\label{sec:fullspectrum}

To extend the WLS absorption length measurement from a single wavelength to a wavelength-dependent spectrum, we use the absorption spectrum $S(\lambda)$ of the EJ-160I core material shown in figure~\ref{fig:spectra}.
The normalized absorption spectrum $S(\lambda)$ defines the relative spectral dependence of the absorption strength, while the absolute scale is determined by the direct measurement at 405.7\,nm.
Expressing the WLS absorption length as the inverse of the corresponding absorption coefficient, we write
\begin{equation}
    L_{\mathrm{WLS}}(\lambda)=\frac{1}{c\,S(\lambda)},
    \label{eq:LWLS}
\end{equation}

where $c$ is a normalization constant. The  constant $c$ is fixed by the measured value of
$L_{\mathrm{WLS}}(405.7\,\mathrm{nm})$, giving
\begin{equation}
    L_{\mathrm{WLS}}(\lambda)
    =
    L_{\mathrm{WLS}}(405.7\,\mathrm{nm})
    \frac{S(405.7\,\mathrm{nm})}{S(\lambda)}.
    \label{eq:reduced}
\end{equation}

\begin{figure}[htbp]
  \centering
  \includegraphics[width=0.9\textwidth]{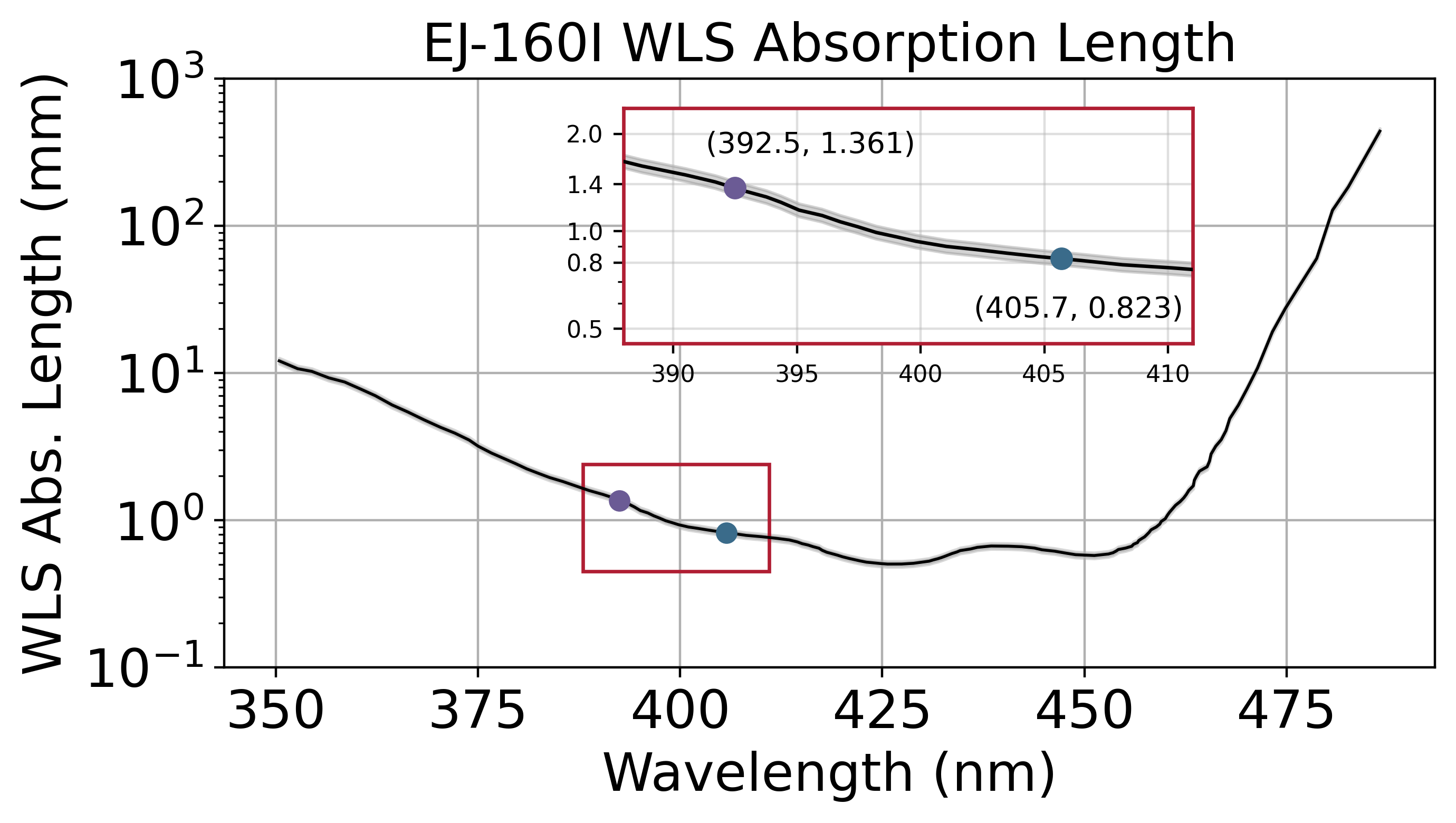}
 \caption{Reconstructed WLS absorption length spectrum of the EJ-160I fiber core, obtained from eq.~(3.4). The inset shows a zoom of the region enclosed by the crimson box, containing the two wavelengths at which the WLS absorption length was measured. In the inset, the blue point indicates the measurement at 405.7~nm ($L_{\mathrm{WLS}} = 0.823~\mathrm{mm}$) used to fix the normalization, and the purple point indicates the reconstructed WLS absorption length at 392.5~nm evaluated from eq.~(3.4), corresponding to the wavelength of the independent measurement discussed in section~\ref{sec:consistency}. The gray band indicates the $\pm1\sigma$ uncertainty.}
  \label{fig:wlsspectrum}
\end{figure}

Applying eq.~\eqref{eq:reduced} to the absorption spectrum $S(\lambda)$ yields the WLS absorption length spectrum shown in figure~\ref{fig:wlsspectrum}.
In this figure, the purple point denotes the WLS absorption length at 392.5\,nm as evaluated from the reconstructed spectrum (eq.~\eqref{eq:reduced}), marking the wavelength at which an independent measurement was subsequently performed for validation (section~\ref{sec:consistency}).


\subsection{WLS absorption length at 392.5 nm}
\label{sec:consistency}

\begin{figure}[h!]
  \centering
  \includegraphics[width=0.8\textwidth]{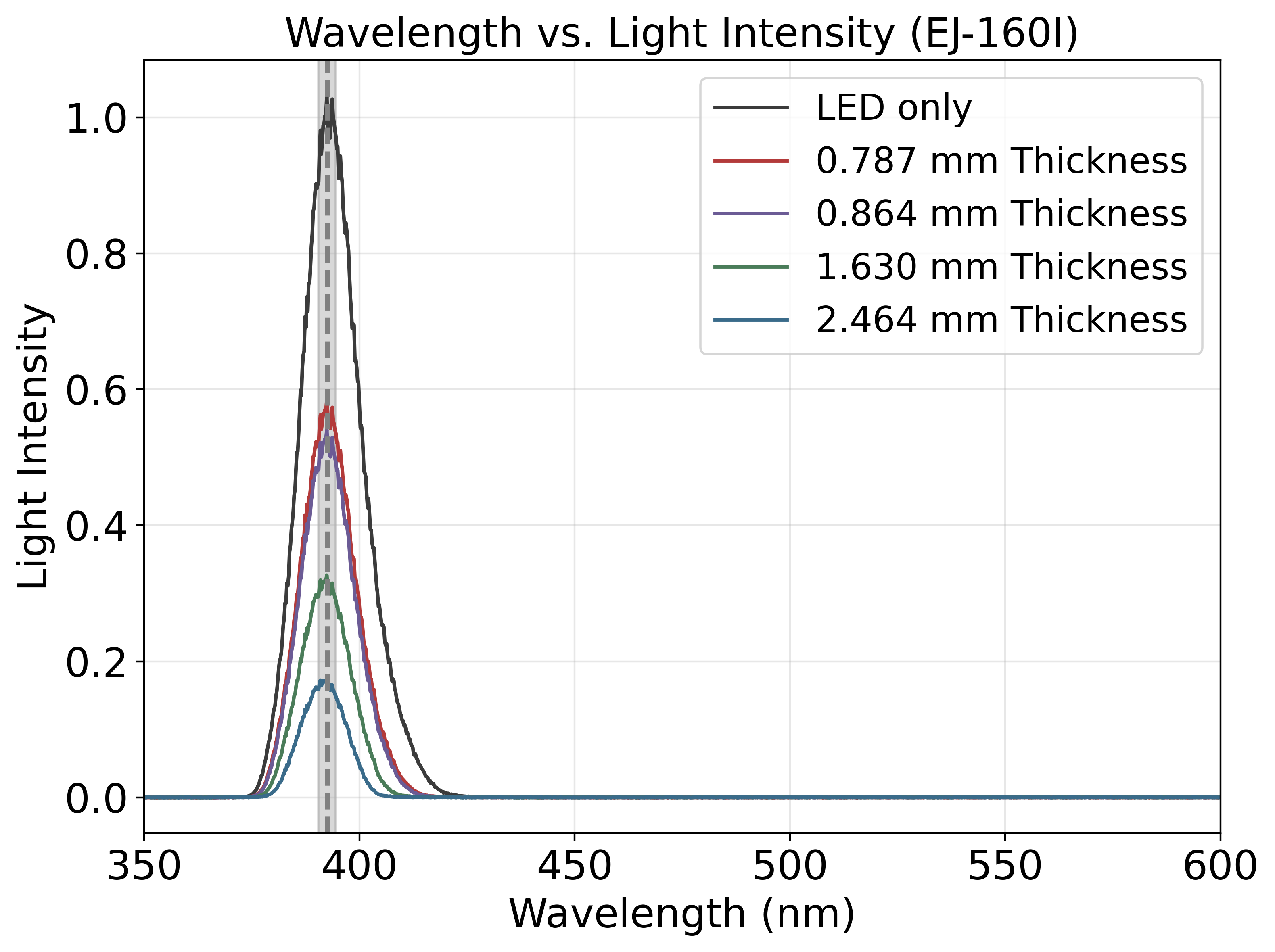}
    \caption{Normalized incident LED intensity and transmitted-light spectra through fiber-core samples with four different thicknesses, corrected for the transmission of the light-guide cable (figure~\ref{fig:xsr}). The shaded gray band indicates the 390.5--394.5~nm window centered on the LED emission peak near 392.5~nm.}
  \label{fig:spectra394}
\end{figure}

\begin{figure}[h!]
  \centering
  \includegraphics[width=0.85\textwidth]{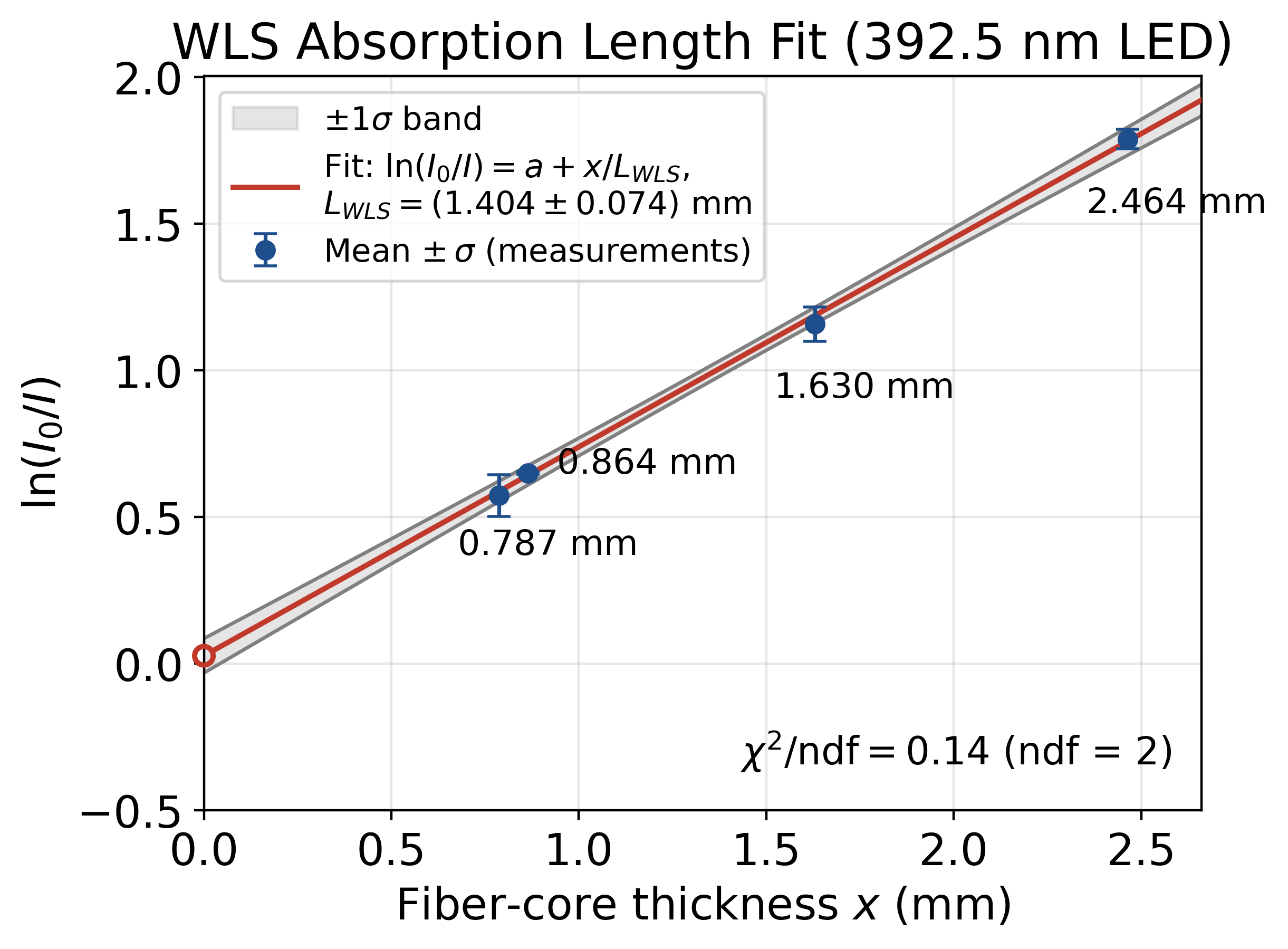}
  \caption{Extraction of the WLS absorption length at 392.5\,nm from the thickness dependence of light attenuation. The intensity ratio $\ln(I_0/I)$ is plotted against fiber-core thickness $x$ for the 392.5~nm LED. The solid red line shows the linear fit $\ln(I_0/I) = a + x/L_{WLS}$ for $x > 0$, yielding $L_{WLS} = (1.404 \pm 0.074)$\,mm. The shaded gray band denotes the corresponding $\pm 1\sigma$ uncertainty band of the fit.}
  \label{fig:fit394}
\end{figure}

To corroborate the reconstructed spectrum, the WLS absorption length was independently measured at 392.5\,nm.
The same transmission-measurement procedure described in sections~\ref{sec:wls_absorption_405} and~\ref{sec:fullspectrum} was repeated using an LED with an emission peak near 392.5~nm; the resulting spectra are shown in figure~\ref{fig:spectra394}.
The transmitted-light intensity at 392.5~nm was extracted for each configuration, and the linear relation between $\ln(I_0/I)$ and the fiber-core thickness is shown in figure~\ref{fig:fit394}.
The inverse of the fitted slope gives
$L_{\mathrm{WLS}}(392.5~\mathrm{nm}) = (1.404 \pm 0.074)$~mm,
with a corresponding fitted optical-coupling factor of
$A = 0.974 \pm 0.057$.
This value differs from that predicted by the reconstructed spectrum by 3.16\%,
corresponding to $0.43\sigma$.

\begin{figure}[h!]
    \centering
    \includegraphics[width=0.7\linewidth]{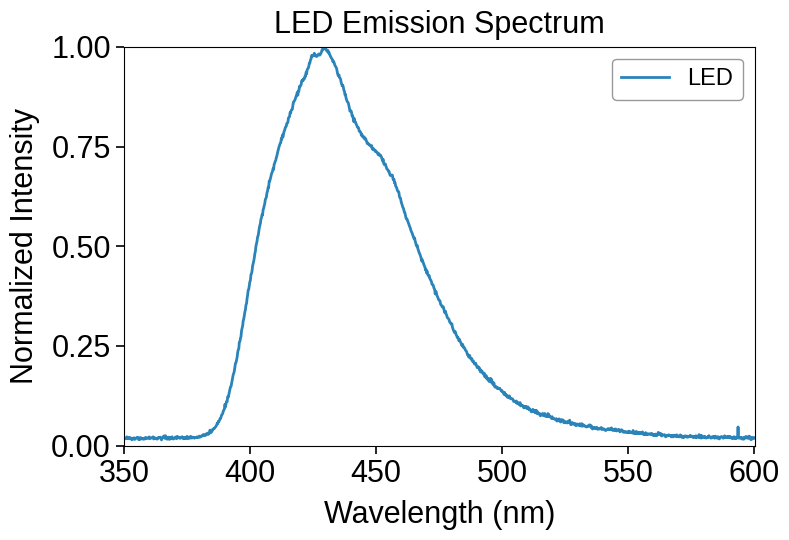}
    \caption{Normalized emission spectrum of the blue LED used in the reference EJ-160I attenuation length measurement and in the corresponding Geant4-based optical Monte Carlo simulation.
    This figure is adapted from~\cite{bae2025optical}.}
    \label{fig:LED}
\end{figure}

\begin{figure}[h!]
    \centering
    \includegraphics[width=0.95\linewidth]{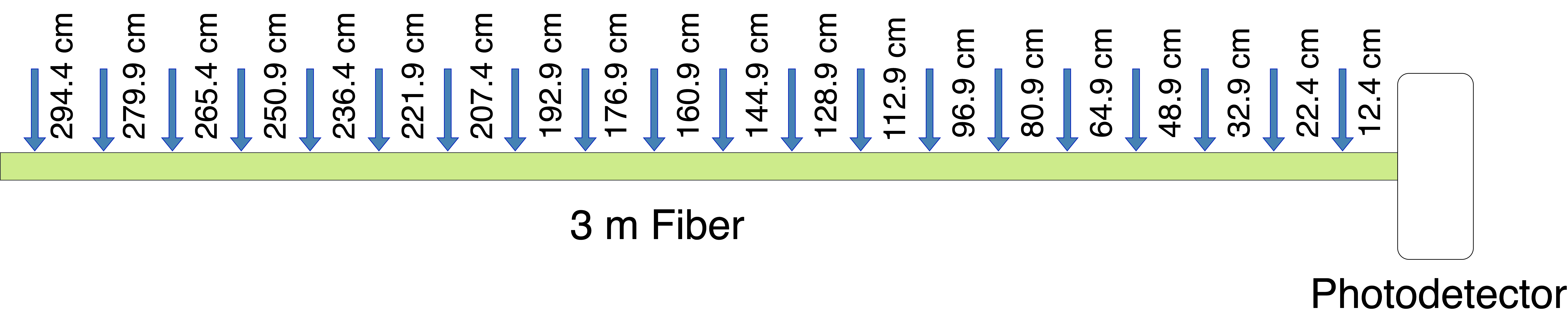}
    \caption{Schematic illustration of the Geant4 optical Monte Carlo geometry used to reproduce the reference EJ-160I attenuation length measurement. Photons were injected perpendicularly onto a straight 3\,m fiber at twenty source positions, and the transmitted light was collected at the photodetector.}
    \label{fig:simulation_setup}
\end{figure}

\section{Monte Carlo Simulation}

\subsection{Comparison with the measured attenuation profile}

As discussed in section~\ref{sec:Introduction}, the meter-scale attenuation length and the millimeter-scale WLS absorption length are implemented in Geant4 optical simulations as separate material properties, \texttt{ABSLENGTH} and \texttt{WLSABSLENGTH}.
To evaluate the reconstructed WLS absorption length spectrum in a fiber-scale optical model, we implemented the spectrum as the \texttt{WLSABSLENGTH} input in a Geant4-based optical Monte Carlo simulation; here, the attenuation length of EJ-160I was taken to be 4.0\,m as the \texttt{ABSLENGTH} input, based on our earlier measurement~\cite{bae2025optical}.
The simulated light attenuation was then compared with the previously reported EJ-160I light attenuation measurement~\cite{bae2025optical}.

In the earlier study~\cite{bae2025optical}, an approximately 3\,m-long EJ-160I fiber was laid in a spiral groove and optically coupled at one end to the same spectrophotometer employed here~\cite{Ocean-Optics}.
A blue LED with a peak emission wavelength of approximately 430\,nm (figure~\ref{fig:LED}) was used to illuminate the fiber from the side at twenty discrete positions corresponding to propagation distances of 0.124--2.944\,m from the spectrophotometer input.
The measured light intensity as a function of source position was used to characterize the meter-scale attenuation length of the fiber.

In the Geant4-based optical Monte Carlo simulation, the simulation geometry was constructed to approximate the experimental attenuation setup while simplifying the fiber layout.
The fiber was modeled in air as a straight 3\,m-long EJ-160I fiber with a
\(1\,\mathrm{mm}\times1\,\mathrm{mm}\) square cross section and a
0.04\,mm-thick PMMA cladding, as shown in figure~\ref{fig:simulation_setup}, rather than the spiral configuration used in the reference measurement~\cite{bae2025optical}.
At each of the twenty source positions, spanning propagation distances from 0.124\,m to 2.944\,m, \(10^5\) photons were generated with a wavelength distribution matching the measured emission spectrum of the same blue LED used in the reference measurement~\cite{bae2025optical}, shown in figure~\ref{fig:LED}, and were incident perpendicularly on the fiber surface.

\begin{figure}[h!]
    \centering
    \includegraphics[width=0.9\linewidth]{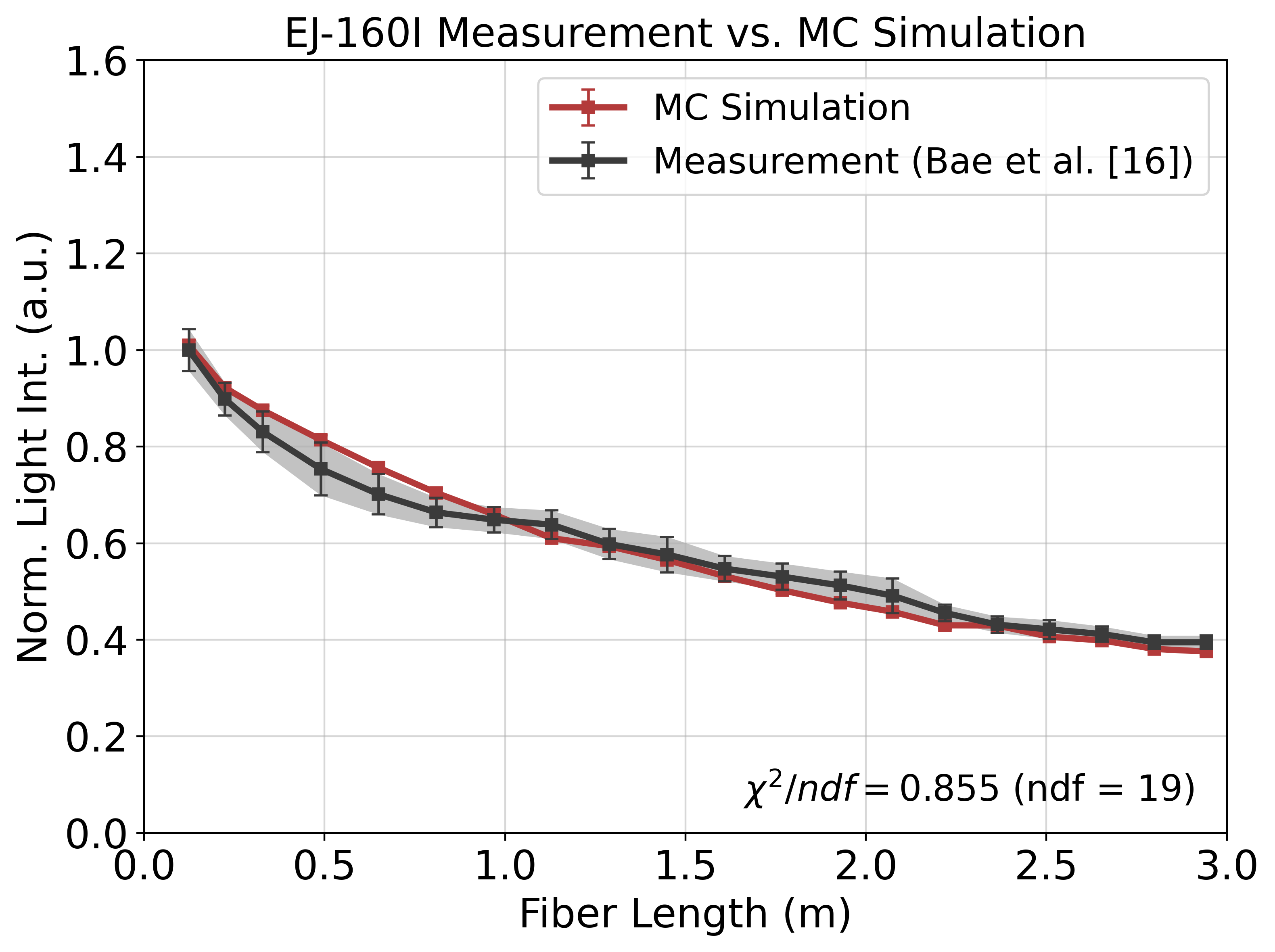}
    \caption{Normalized light intensity as a function of fiber length for EJ-160I, comparing the Geant4-based Monte Carlo (MC) simulation with the measurements reported in \cite{bae2025optical}. The simulated and measured profiles are normalized to a common mean value to compare their relative attenuation with fiber length. The shaded band represents the measurement uncertainty, including sample-to-sample variation.}   
    \label{fig:simulation}
\end{figure}

The resulting simulated light attenuation profile is shown in figure~\ref{fig:simulation}, together with the measured light attenuation profile reported in the previous study~\cite{bae2025optical}.
The agreement between the simulation and measurement is quantified by $\chi^2/\mathrm{ndf} = 0.855~(\mathrm{ndf = 19})$, indicating that the simulation reproduces the overall
attenuation trend observed in the measurement.

In measurements, the observed light output may vary between fibers even for the same fiber model due to imperfections or contamination on the fiber surface, such as microscopic abrasions, dust, and residues, as well as differences in the boundary conditions arising from contact with other materials in the experimental setup~\cite{Amos1990:fiber-cladding_light,Achenbach:fiber-cladding_light}.
These nontrivial imperfection effects are not explicitly modeled in the simulation.


\subsection{Role of the WLS absorption length}

\texttt{WLSABSLENGTH}, the Geant4 input parameter, sets the characteristic distance over which an
incident photon is absorbed by the WLS fluor prior to re-emission, and therefore
determines the fraction of incident light that is wavelength-shifted and guided,
rather than the spectral distribution of the re-emitted light.
To illustrate this, two complementary aspects of the simulation were examined:
the total detected light output and the detected emission spectrum.

\begin{figure}[h!]
    \centering
    \includegraphics[width=0.8\linewidth]{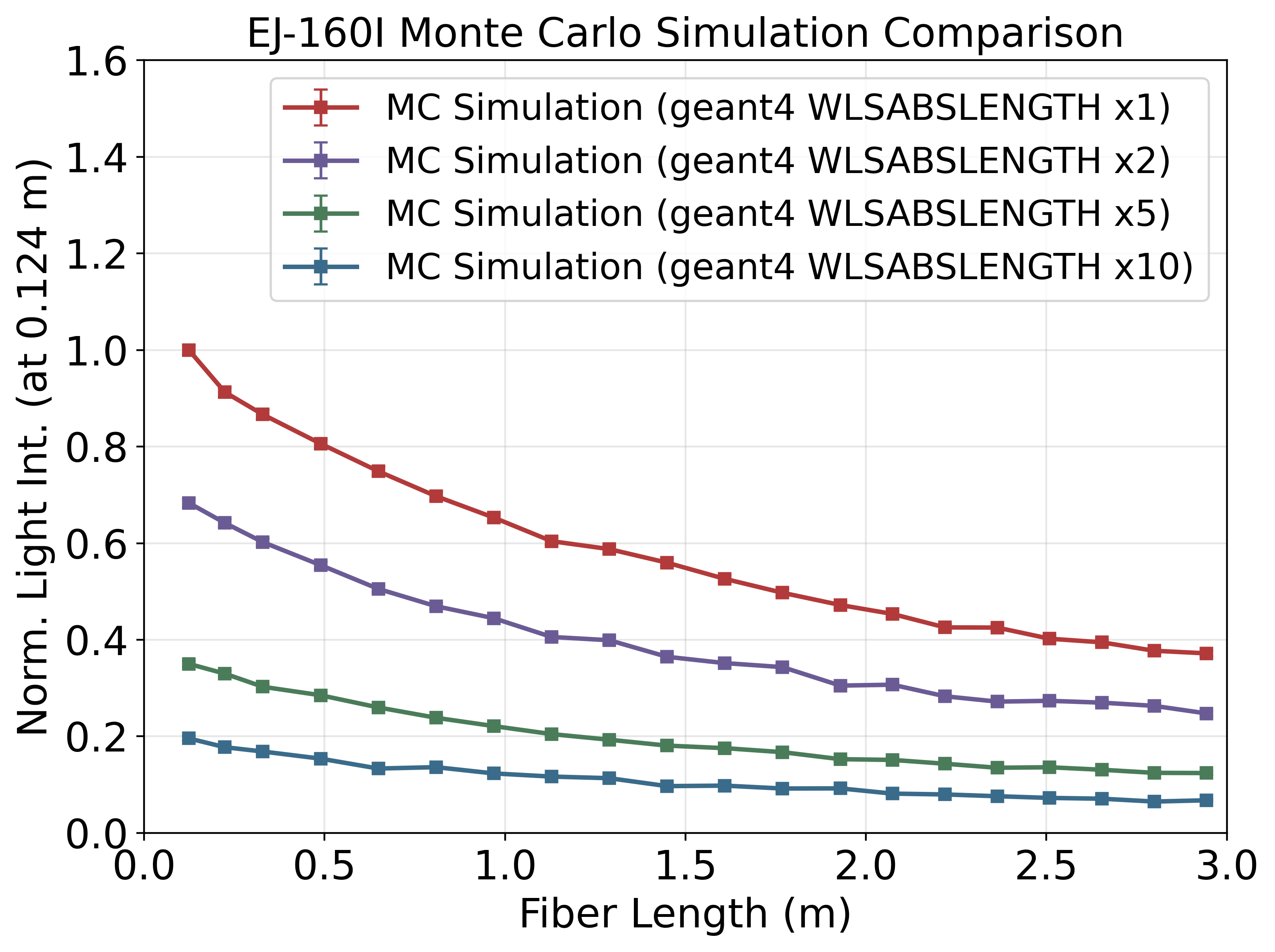}
    \caption{Effect of the WLS absorption length on the simulated light output. The Geant4 \texttt{WLSABSLENGTH} material property~\cite{geant4} was scaled uniformly to evaluate this dependence. 
    The reconstructed WLS absorption length spectrum from figure~\ref{fig:wlsspectrum} was used as the $\times1$ input, together with the same spectrum uniformly scaled by factors of 2, 5, and 10.
    The $\times1$ case is identical to the Geant4 simulation curve shown in figure~\ref{fig:simulation}.
    All curves are normalized to the $\times1$ result at the shortest distance, 0.124\,m. A larger WLS absorption length lowers the overall light output.}
    \label{fig:wls_scaling}
\end{figure}

Figure~\ref{fig:wls_scaling} shows the simulated light output as a function of
fiber length obtained with the reconstructed WLS absorption length spectrum and
with the same spectrum uniformly scaled by factors of 2, 5, and 10.
A larger WLS absorption length reduces the fraction of incident photons absorbed
per unit thickness, so the overall detected light decreases accordingly, whereas
the shape of the light attenuation curve is largely preserved.
The distance dependence is instead governed by the meter-scale attenuation length (\texttt{ABSLENGTH}), which was held fixed.

\begin{figure}[h!]
    \centering
    \includegraphics[width=0.8\linewidth]{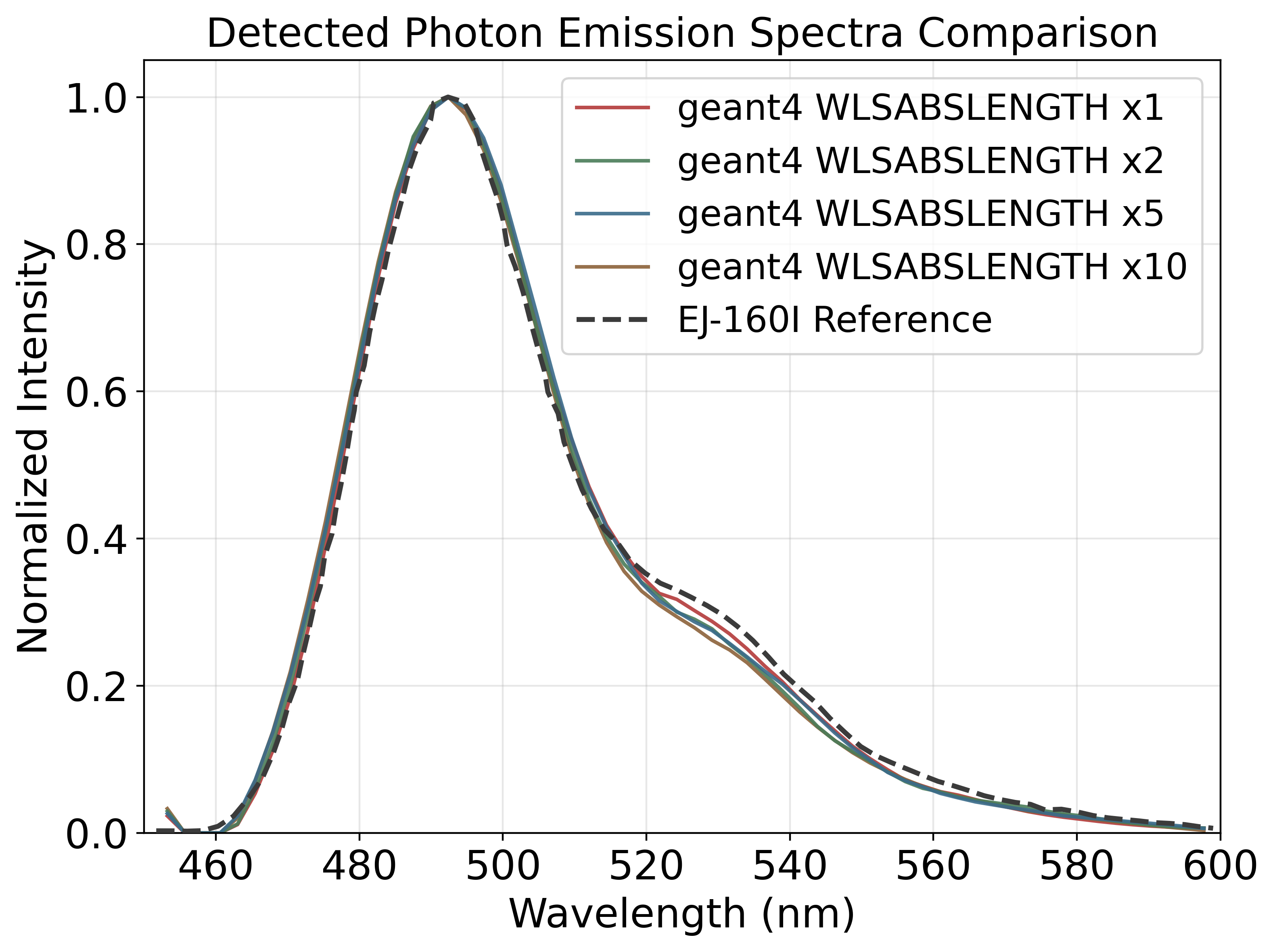}
    \caption{Effect of the WLS absorption length on the simulated light emission spectrum.
    Simulated photon emission spectra (solid lines), obtained from Geant4 optical Monte Carlo simulations with photons injected perpendicularly 1~mm from the photodetector, are compared with the reference EJ-160I emission spectrum (dashed line) shown in figure~\ref{fig:spectra}.
    The WLS absorption length was scaled by factors of 2, 5, and 10 relative to the spectrum in figure~\ref{fig:wlsspectrum}; to maintain comparable statistical precision, the number of generated photons was increased by the same factors ($\times 2$, $\times 5$, and $\times 10$) for each configuration.
    The simulated spectrum reproduces the reference across the emission band regardless of the scale of WLS absorption length.}
    \label{fig:emission_compare}
\end{figure}

Figure~\ref{fig:emission_compare} compares the detected photon emission spectrum
from the simulation with the reference EJ-160I emission spectrum.
The spectrum of detected photons reproduces the reference across the full emission band.
Because the re-emission spectrum is fixed by the emission-spectrum input and is
independent of \texttt{WLSABSLENGTH}, changing the scale of WLS absorption length alters
only the light output and leaves the emission spectrum unchanged.
This demonstrates our expectation that the WLS absorption length determines the fraction of incident light absorbed by the WLS fluors in the fiber core, while being decoupled from the emission spectral shape.

\section{Conclusion}

This study presents direct measurements of the wavelength-shifting (WLS) absorption length of the EJ-160I fiber core at two wavelengths.
The measured values were \((0.823 \pm 0.038)\,\mathrm{mm}\) at \(405.7\,\mathrm{nm}\) and \((1.404 \pm 0.074)\,\mathrm{mm}\) at \(392.5\,\mathrm{nm}\), with the latter differing from the reconstructed spectrum by 3.16\%, corresponding to \(0.43\sigma\).
The reconstructed spectrum was implemented as the \texttt{WLSABSLENGTH} input in a Geant4-based optical Monte Carlo simulation of light attenuation in the EJ-160I fiber.
The simulation shows that the WLS absorption length specifically affects the overall wavelength-shifted light output while leaving the emission spectrum unchanged.
This transmission-based characterization framework provides a practical method for constructing WLS absorption length spectra for WLS-fiber simulations in future particle and nuclear physics experiments.

\acknowledgments{

We thank Eljen Technology for providing EJ-160I fiber samples. This work was supported in part by the University of Texas at Austin and the U.S. National Science Foundation under grant PHY-2312278.
}
\end{document}